\documentclass{article}
\usepackage{spconf,amsmath,graphicx, tikz, pgfplots, pgfplotstable, subfig}

\graphicspath{{./data/}}
\pgfplotsset{select coords between index/.style={
    x filter/.code={
        \ifnum\coordindex<#1\fi
    }
}, table/search path={data}}
\pgfplotsset{compat=1.16}

\newcommand{\transp}{{\sf T}}
\newcommand{\abs}[1]{\lvert#1\rvert}
\newcommand{\norm}[1]{\left\Vert#1\right\Vert}

\newcommand{\arr}[1]{\mathbf{#1}}
\newcommand{\greekarr}[1]{\boldsymbol{#1}}
\newcommand{\setcurve}[1]{\mathcal{#1}}

\definecolor{blue1}{RGB}{57, 106, 177}
\definecolor{red1}{RGB}{204, 37, 41}

\pgfplotscreateplotcyclelist{my col}{%
blue1,\\%
red1, densely dashed, line width=2pt\\%
}

\title{Towards bandwidth estimation for graph signal reconstruction}

\name{Ajinkya Jayawant, Antonio Ortega}

\address{Department of Electrical and Computer Engineering, University of Southern California}

\begin{document}
\maketitle
\begin{abstract}
In numerous graph signal processing applications, data is often missing for a variety of reasons, and predicting the missing data is essential. In this paper, we consider data on graphs modeled as bandlimited graph signals. Predicting or reconstructing the unknown signal values for such a model requires an estimate of the signal bandwidth. In this paper, we address the problem of estimating the reconstruction errors, minimizing which would thereby provide an estimate of the signal bandwidth. In doing so, we design a cross-validation approach needed for stable graph signal reconstruction and propose a method for estimating the reconstruction errors for different choices of signal bandwidth. Using this technique, we are able to estimate the reconstruction error on a variety of real-world graphs.
\end{abstract}

\begin{keywords}
Graph, signal, bandwidth, reconstruction, estimation.
\end{keywords}

\section{Introduction}
\label{sec:intro}
Graphs naturally arise in a number of applications such as data analysis in various types of networks (e.g., sensor, traffic or  social networks).
Information related to the nodes in the network constitutes a graph signal \cite{ortega2021introduction}.
Graph signals in practical scenarios are often incomplete due to a variety of reasons -- sensor failure \cite{lopes2005inferential}, occlusions \cite{guerreiro2002factorization}, measurements outside range of sensors \cite{mott1994climate}. As a result, the values of a graph signal on certain nodes may be unknown to us.
In theory, we can predict missing data on the graph provided that the signal is smooth. This prediction is perfect if signal values are known on enough nodes and the underlying graph signal is bandlimited \cite{chen2015discrete}, \cite{tanaka2020sampling}.
Given the bandwidth of the graph signal, the original signal can be predicted or reconstructed by solving an inverse problem \cite{anis2015efficient}, \cite{jayawant2022practical}.

Note that reconstructing the signal using the bandlimited signal model requires knowledge of the bandwidth of the signal.
Most papers in the graph signal processing literature \cite{chen2015discrete,puy2016random,anis2015efficient,jayawant2022practical} simply assume that this bandwidth is known.
However, in many real scenarios, only the signal values on a limited number of nodes are known, while the bandwidth of the signal is unknown.
To add to this difficulty, in reality signals are not bandlimited to a certain maximum frequency because of either (i)  noise in the signal, or  (ii) a mismatch between the chosen graph construction and a hypothetical graph construction for which observed signals would be bandlimited.

Even if the exact signal bandwidth is unknown (or if the signal is not exactly bandlimited),  reconstruction using the bandlimited model of the signal remains useful because it signifies signal smoothness which is a reasonable assumption for many real-life signals like temperature.
This means that it is important to optimize the choice of the bandwidth of the signal for reconstruction, regardless of whether the original signal is bandlimited or not. Since the reconstruction error of the signal usually varies with the choice of the reconstruction bandwidth, the bandwidth which gives us the smallest error over a choice of different bandwidths would be the right choice for reconstruction.
To select a bandwidth in such a way, we need the know reconstruction error for different choices of bandwidth. However, we cannot calculate the overall reconstruction error since there are missing signal values. Thus, we need an estimate of the actual reconstruction error.

Our main contribution is to formulate the problem of selecting a reconstruction bandwidth from data, without knowledge of the actual graph signal bandwidth, a problem as yet not considered in the graph signal sampling and reconstruction literature (see \cite{tanaka2020sampling} for a review). We propose a solution  that uses a novel cross-validation methodology based on graph signal sampling concepts. Specifically, we solve the problem of estimating the reconstruction error which is essential to select a reconstruction bandwidth.

In a standard cross-validation setting, multiple random subsets are used to validate parameter choices. We show that using random subsets for our problem can result in ill-conditioned reconstruction operators and propose a technique that mitigates the effects of ill-conditioning by giving different importance to each random subset. This approach significantly improves the error estimation,
and our proposed method estimates the squared reconstruction error with good accuracy for a wide variety of both  synthetic and real-life graphs and
  signals.

\section{Problem formulation}
\label{sec:prob_form}

\subsection{Notation}
\label{sec:typestyle}

For a graph with $n$ nodes,
the $ij$ entry $w_{ij}$ of the $n\times n$ weighted adjacency matrix $\arr{A}$ is edge weight between the $i^\text{th}$ and $j^\text{th}$ nodes, with
$w_{ii} =0$.
The degree matrix $\mathbf{D}$  is a diagonal matrix with entries $d_{ii}= \sum_j w_{ij}$.
The combinatorial Laplacian is given by $\mathbf{L} = \mathbf{D} - \mathbf{A}$, with its corresponding eigendecomposition defined as  $\mathbf{L} = \mathbf{U} \Sigma \mathbf{U}^\transp$ since the Laplacian matrix is symmetric and positive semidefinite.

\begin{table}[t]
    \centering
    \caption{Notation}
    \begin{tabular}{| c | c |}\hline
        Notation & Description \\ \hline
        $\mathbf{A}_{\mathcal{XY}}$ & Submatrix of $\mathbf{A} $ indexed by $\mathcal{X}$(rows)\\
         &  and $\mathcal{Y}$(columns)\\ \hline
        $\mathbf{A}_{\mathcal{X}}$ & $\mathbf{A}_{\mathcal{:,X}}$, columns of $\mathbf{A}$ indexed by $\setcurve{X}$\\ \hline
        $\mathbf{x}_{\mathcal{X}}$ or $\mathbf{x}(\mathcal{X})$ & Subset of $\mathbf{x}$ indexed by $\mathcal{X}$\\ \hline
    \end{tabular}
    \label{tab:papNote}
\end{table}

The column vectors of $\arr{U}$ provide a basis with frequency interpretation for graph signals \cite{shuman2013emerging}, and  the operator $\mathbf{U}^\transp$ is usually called the graph Fourier transform (GFT).
The eigenvectors $\mathbf{u}_i$ of $\mathbf{L}$ associated with larger eigenvalues $\lambda_i$ correspond to higher frequencies, and the ones associated with lower eigenvalues correspond to lower frequencies \cite{ortega2021introduction}, with the following convention for the indexing of the eigenvalues: $\lambda_1 \leq \lambda_2 \leq \cdots \leq\lambda_n$.
We  denote $\arr{x}$ a vector, $\arr{X}$ a matrix, and $\mathcal{X}$  a set.
We also follow conventions from Table \ref{tab:papNote}.

\subsection{Graph signals and reconstruction}
We consider $n$-dimensional scalar real-valued signal $\mathbf{x}$ on the vertex set $\mathcal{V}$. In line with the problem of missing data, we assume that only a part of this signal is known, corresponding to a subset of vertices,  $\mathcal{S}\subseteq \mathcal{V}$. We denote $\mathbf{x}_{\mathcal{S}}$ and  $\mathbf{x}_{\mathcal{S}^c}$ the known and unknown signals, respectively, where $\setcurve{S}^c$ is the complement of $\setcurve{S}$.
Estimating $\mathbf{x}_{\mathcal{S}^c}$ from $\mathbf{x}_{\mathcal{S}}$ is the graph signal reconstruction problem \cite{pesenson2008sampling}. We denote the reconstructed unknown signal as $\hat{\arr{x}}_{\setcurve{S}^c}$, and quantify its closeness with the original signal, $\mathbf{x}_{\mathcal{S}^c}$, using the $\ell_2$ norm $\norm{\mathbf{x}_{\mathcal{S}^c} - \hat{\arr{x}}_{\setcurve{S}^c}}_2^2$. However, for signal reconstruction a signal model needs to be chosen. A popular choice for a smooth graph signal is the bandlimited signal model \cite{anis2015efficient}.

\subsection{Signal model}
\label{sec:sigMod}

In this paper we consider bandlimited signals defined as $\arr{f} = \arr{U}_{\setcurve{R}}\greekarr{\alpha}$, where $\mathcal{R}$ is the set $\{1, \cdots, r\}$, and $\boldsymbol{\alpha}$ is an $r$-dimensional vector. We call $r$  the bandwidth of the signal.
However, graph signals are rarely exactly bandlimited, so in this paper we consider the following more realistic model of a bandlimited signal with added noise:
\begin{equation}
\arr{x} = \arr{U}_{\setcurve{R}}\greekarr{\alpha} + \arr{n},
\label{eq:noiseModel}
\end{equation}
where $\mathbf{n}$ is an $n$-dimensional noise vector.

\subsection{Model selection for reconstruction}
With the signal model in \eqref{eq:noiseModel} and known signal values $\mathbf{x}_{\mathcal{S}}$, the signal on  $\setcurve{S}^c$ can be reconstructed as:
\begin{equation*}
    \hat{\arr{x}}_{\setcurve{S}^c} = \arr{U}_{\setcurve{S}^c\setcurve{R}}(\arr{U}_{\setcurve{SR}}^\transp \arr{U}_{\setcurve{SR}})^{-1}\arr{U}_{\setcurve{SR}}^\transp\arr{x}_{\setcurve{S}}.
\end{equation*}
This is a least squares reconstruction and is typical \cite{anis2015efficient} when the size of the known signal set is larger than the signal bandwidth used for reconstruction, $\abs{\setcurve{S}} > r$, which is the setting we consider in this paper. Note that this reconstruction requires the signal bandwidth $r$ to be known, regardless of whether the signal is bandlimited or bandlimited with additional noise. Most reconstruction algorithms assume that this bandwidth is known \cite{chen2015discrete, puy2016random, wang2015local}. However, fundamentally this is a model selection problem where an appropriate bandlimited signal model with a fixed bandwidth $r$ needs to be chosen.

\subsection{Bandwidth selection through reconstruction errors}
Although the goal of model selection for signal reconstruction is to choose $r$, the signal itself might not be bandlimited. As a result, there may not be any prior for signal bandwidth. However, our primary goal is to minimize the reconstruction error: $E_{\setcurve{S}^c} = \norm{\mathbf{x}_{\mathcal{S}^c} - \hat{\arr{x}}_{\setcurve{S}^c}}^2$, where the estimate $\hat{\arr{x}}_{\setcurve{S}^c}$ is a function of $r$, and so is $E_{\setcurve{S}^c}$. To select $r$ we propose a minimization of $\norm{\hat{\arr{x}}_{\setcurve{S}^c} - \arr{x}_{\setcurve{S}^c}}^2$ over a set of possible values of $r$, so that whichever bandwidth $r$ minimized the error will be used as the reconstruction bandwidth, $r^* = \min_r E_{\setcurve{S}^c}$.

However, minimizing $\norm{\hat{\arr{x}}_{\setcurve{S}^c} - \arr{x}_{\setcurve{S}^c}}^2$ is impossible without knowing $\arr{x}_{\setcurve{S}^c}$. For that reason, we propose estimating the error $\norm{\hat{\arr{x}}_{\setcurve{S}^c} - \arr{x}_{\setcurve{S}^c}}^2$ for different values of $r$ using the known signal values, $\arr{x}_\setcurve{S}$. We limit the scope of this paper to estimating this reconstruction error and leave the bandwidth selection for future work. Towards that end, we want an estimate, $\hat{E}_{\setcurve{S}^c}$, of the  reconstruction error $E_{\setcurve{S}^c}$ for different values of $r$, with $\abs{E_{\setcurve{S}^c} -\hat{E}_{\setcurve{S}^c}}$ as small as possible.

\section{Cross-validation theory for graph signals}
\label{sec:theory}

In order to accurately estimate the reconstruction error as a function of the signal bandwidth $r$, it is essential to analyze in more detail the error with respect to subset selection on the set of graph vertices.

\subsection{Conventional error estimation and shortcomings}
\label{sec:ill_cond_prob}
The reconstruction error, $\arr{e}(\setcurve{S}^c)$, measured over the unknown nodes is the following:
\begin{equation*}
    \arr{e}(\setcurve{S}^c) = \arr{x}_{\setcurve{S}^c} - \hat{\arr{x}}_{\setcurve{S}^c} = \arr{x}_{\setcurve{S}^c} - \arr{U}_{\setcurve{S}^c\setcurve{R}} (\arr{U}_{\setcurve{S}\setcurve{R}}^\transp \arr{U}_{\setcurve{S}\setcurve{R}})^{-1}\arr{U}_{\setcurve{S}\setcurve{R}}^\transp\arr{x}_{\setcurve{S}}.
\end{equation*}
  To estimate this error we could split the set $\setcurve{S}$ further into the sets $\{\setcurve{S}_1, \setcurve{S}_1^c\}$,$\cdots$, $\{\setcurve{S}_k, \setcurve{S}_k^c\}$  such that $\setcurve{S}_i \cup \setcurve{S}_i^c = \setcurve{S}$ for $i \in \{1, \cdots, k\}$ and estimate
\begin{equation*}
    \arr{e}(\setcurve{S}_i^c) = \arr{x}_{\setcurve{S}_i^c} - \arr{U}_{\setcurve{S}_i^c\setcurve{R}} (\arr{U}_{\setcurve{S}_i\setcurve{R}}^\transp \arr{U}_{\setcurve{S}_i\setcurve{R}})^{-1}\arr{U}_{\setcurve{S}_i\setcurve{R}}^\transp\arr{x}_{\setcurve{S}_i}
\end{equation*}
and use the estimate $\hat{E}_{\setcurve{S}^c} = \sum_{i\in\{1, \cdots, k\}} \norm{\arr{e}(\setcurve{S}_i^c)}^2/k$.
This would be equivalent to using the standard cross-validation approach that is typical in linear model selection \cite{shao1993linear}.

The bandlimited component of the signal has no effect on either $\arr{e}(\setcurve{S}^c)$ or $\arr{e}(\setcurve{S}_i^c)$.
Suppose that the noise vector has some representation, $\arr{n} = \arr{U}_{\setcurve{R}}\greekarr{\gamma} + \arr{U}_{\setcurve{R}^c}\greekarr{\beta}$, we can conveniently separate the bandlimited and non-bandlimited components of the signal using the following representation: $\arr{x} = \arr{U}_{\setcurve{R}}\greekarr{\alpha}' + \arr{U}_{\setcurve{R}^c}\greekarr{\beta}$, where $\greekarr{\alpha}' = \greekarr{\alpha} + \greekarr{\gamma}$.

Using the new representation of the signal, with the following simplification of notation,
\begin{align*}
    \arr{M} &= \arr{U}_{\setcurve{S}^c\setcurve{R}^c} - \arr{U}_{\setcurve{S}^c\setcurve{R}} (\arr{U}_{\setcurve{S}\setcurve{R}}^\transp \arr{U}_{\setcurve{S}\setcurve{R}})^{-1}\arr{U}_{\setcurve{S}\setcurve{R}}^\transp\arr{U}_{\setcurve{S}\setcurve{R}^c}\\
    \arr{M}_i &= \arr{U}_{\setcurve{S}_i^c\setcurve{R}^c} - \arr{U}_{\setcurve{S}_i^c\setcurve{R}} (\arr{U}_{\setcurve{S}_i\setcurve{R}}^\transp \arr{U}_{\setcurve{S}_i\setcurve{R}})^{-1}\arr{U}_{\setcurve{S}_i\setcurve{R}}^\transp\arr{U}_{\setcurve{S}_i\setcurve{R}^c}
\end{align*}
our errors are
\begin{equation*}
    \arr{e}(\setcurve{S}^c) = \arr{M}\greekarr{\beta},\quad \arr{e}(\setcurve{S}_i^c) = \arr{M}_i\greekarr{\beta} \quad i\in\{1,\cdots,k\}.
\end{equation*}
The matrices $\arr{M}$ and $\arr{M}_i$ are what mainly differentiate the errors $\arr{e}(\setcurve{S}^c)$ and $\arr{e}(\setcurve{S}_i^c)$. Because the subsets $\setcurve{S}_i$ are selected randomly, $\arr{M}_i$ can be ill-conditioned although $\arr{M}$ is well-conditioned. This ill-conditioning often causes the estimate of the cross-validation error to be orders of magnitude higher than the actual error.

\begin{figure}
    \centering
    \includegraphics[width=0.6\linewidth]{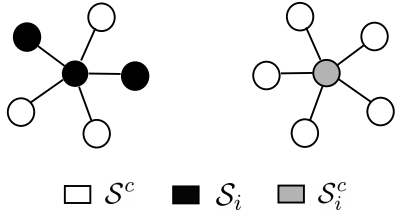}
    \caption{Ill-conditioning scenario for a cross-validation subset.}
    \label{fig:subset_ill}
\end{figure}

Intuitively, this can happen in cases where $\setcurve{S}$ is well connected to $\setcurve{S}^c$ but $\setcurve{S}_i$ is poorly connected to $\setcurve{S}_i^c$. See Fig. \ref{fig:subset_ill} for a toy example where all vertices in the graph are within one hop of $\setcurve{S}$, but $\setcurve{S}_i$ is disconnected from $\setcurve{S}_i^c$. Trying to reconstruct the signal on $\setcurve{S}_i^c$ using the known signal values on $\setcurve{S}_i$ is impossible, and that would be reflected as an ill-conditioned $\arr{M}_i$. Although this is a corner case example, it can be generalized to similar situations, either less or more extreme, arising when using random graphs subsets for reconstruction.

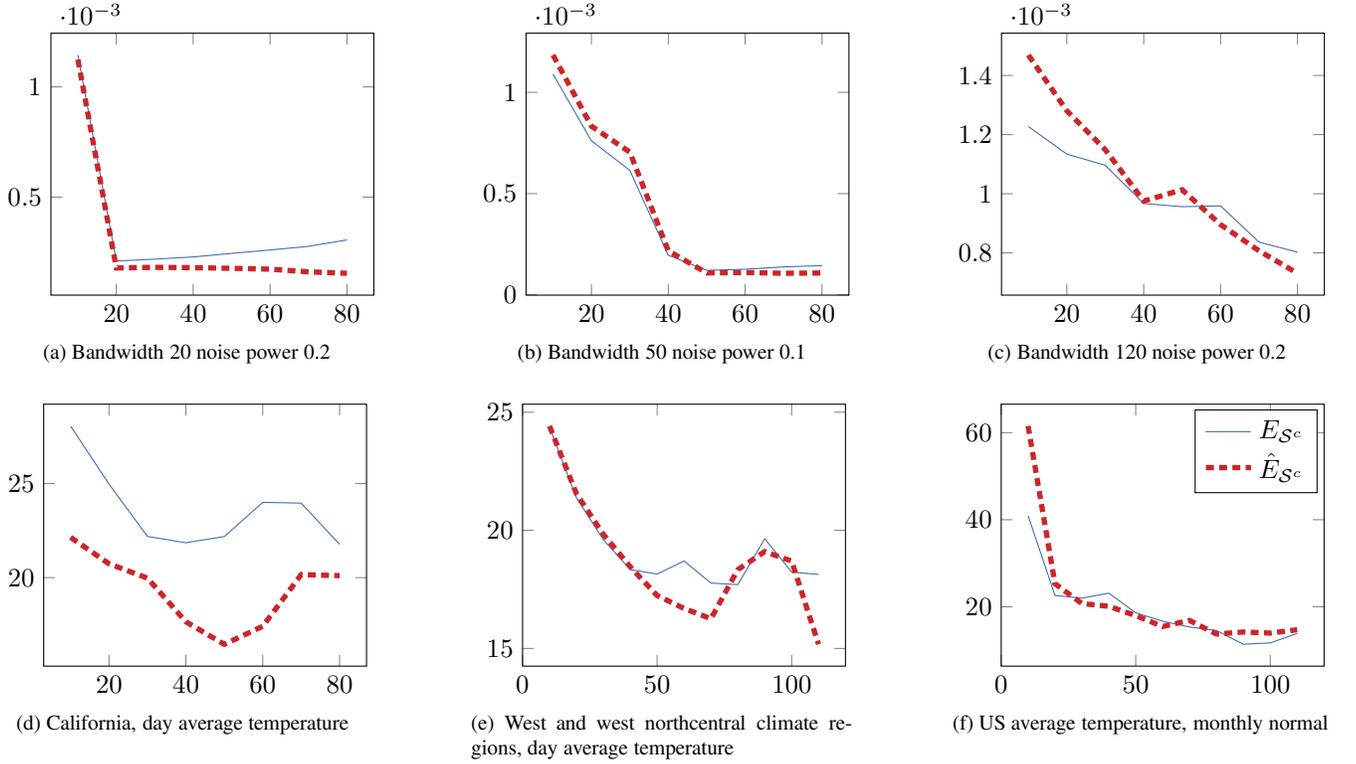
\begin{figure*}[th]
\subfloat[Bandwidth 20 noise power 0.2]{
\pgfplotstableread{data/synthetic_different_noise/2022_10_24_16_55_53/recon_valid_normalized.txt}\actualErrorDat
\pgfplotstableread{data/synthetic_different_noise/2022_10_24_16_55_53/recon_cross_normalized.txt}\cvErrorDat
\begin{tikzpicture}
\centering
  \begin{axis}[
  	width = 0.33\linewidth, no marks,
  	cycle list name=my col]
  	\addplot table [
x={create col/set list={10,20,30,40,50,60,70,80,90,100,110}},
y index=0
] {\actualErrorDat};
\addplot table [
x={create col/set list={10,20,30,40,50,60,70,80,90,100,110}},
y index=0
] {\cvErrorDat};
  \end{axis}
  \label{fig:bw20npp2}
\end{tikzpicture}
}\hfill
\subfloat[Bandwidth 50 noise power 0.1]{
\pgfplotstableread{data/synthetic_different_noise/2022_10_24_16_40_33/recon_valid_normalized.txt}\actualErrorDat
\pgfplotstableread{data/synthetic_different_noise/2022_10_24_16_40_33/recon_cross_normalized.txt}\cvErrorDat
\begin{tikzpicture}
\centering
  \begin{axis}[
  	width = 0.33\linewidth, no marks,
  	cycle list name=my col]
  	\addplot table [
x={create col/set list={10,20,30,40,50,60,70,80,90,100,110}},
y index=0
] {\actualErrorDat};
\addplot table [
x={create col/set list={10,20,30,40,50,60,70,80,90,100,110}},
y index=0
] {\cvErrorDat};
  \end{axis}
\end{tikzpicture}
}\hfill
\subfloat[Bandwidth 120 noise power 0.2]{
\pgfplotstableread{data/synthetic_different_noise/2022_10_24_17_3_34/recon_valid_normalized.txt}\actualErrorDat
\pgfplotstableread{data/synthetic_different_noise/2022_10_24_17_3_34/recon_cross_normalized.txt}\cvErrorDat
\begin{tikzpicture}
\centering
  \begin{axis}[
  	width = 0.33\linewidth, no marks,
  	cycle list name=my col]
  	\addplot table [
x={create col/set list={10,20,30,40,50,60,70,80,90,100,110}},
y index=0
] {\actualErrorDat};
\addplot table [
x={create col/set list={10,20,30,40,50,60,70,80,90,100,110}},
y index=0
] {\cvErrorDat};
  \end{axis}
\end{tikzpicture}
}\hfill\\
\subfloat[California, day average temperature]{
\pgfplotstableread{real_data_experiments/2022_10_24_15_16_34/recon_valid_normalized.txt}\actualErrorDat
\pgfplotstableread{real_data_experiments/2022_10_24_15_16_34/recon_cross_normalized.txt}\cvErrorDat
\begin{tikzpicture}
\centering
  \begin{axis}[
  	width = 0.33\linewidth, no marks,
  	cycle list name=my col]
  	\addplot table [
x={create col/set list={10,20,30,40,50,60,70,80,90,100,110}},
y index=0
] {\actualErrorDat};
\addplot table [
x={create col/set list={10,20,30,40,50,60,70,80,90,100,110}},
y index=0
] {\cvErrorDat};
  \end{axis}
\end{tikzpicture}
}\hfill
\subfloat[West and west northcentral climate regions, day average temperature]{
\pgfplotstableread{data/real_data_experiments/2022_10_24_15_58_5/recon_valid_normalized.txt}\actualErrorDat
\pgfplotstableread{data/real_data_experiments/2022_10_24_15_58_5/recon_cross_normalized.txt}\cvErrorDat
\begin{tikzpicture}
\centering
  \begin{axis}[
  	width = 0.33\linewidth, no marks,
  	cycle list name=my col]
  	\addplot table [
x={create col/set list={10,20,30,40,50,60,70,80,90,100,110}},
y index=0
] {\actualErrorDat};
\addplot table [
x={create col/set list={10,20,30,40,50,60,70,80,90,100,110}},
y index=0
] {\cvErrorDat};
  \end{axis}
\end{tikzpicture}
}\hfill
\subfloat[US average temperature, monthly normal]{
\pgfplotstableread{data/real_data_experiments/2022_10_24_15_42_27/recon_valid_normalized.txt}\actualErrorDat
\pgfplotstableread{data/real_data_experiments/2022_10_24_15_42_27/recon_cross_normalized.txt}\cvErrorDat
\begin{tikzpicture}
\centering
  \begin{axis}[
  	width = 0.33\linewidth, no marks,
  	cycle list name=my col]
  	\addplot table [
x={create col/set list={10,20,30,40,50,60,70,80,90,100,110}},
y index=0
] {\actualErrorDat};
\addplot table [
x={create col/set list={10,20,30,40,50,60,70,80,90,100,110}},
y index=0
] {\cvErrorDat};
\addlegendimage{blue1}
    \addlegendentry{$E_{\setcurve{S}^c}$};
    \addlegendimage{red1, densely dashed, line width=2pt}
    \addlegendentry{$\hat{E}_{\setcurve{S}^c}$};
  \end{axis}
\end{tikzpicture}
}\hfill
\caption{Squared reconstruction errors vs bandwidth for bandlimited signal model. Legend is common for all the plots.}
\label{fig:mainRecon}
\end{figure*}

\subsection{Proposed error estimation}
As we noted in Section \ref{sec:ill_cond_prob}, averaging of the error over random subsets may lead to blowing up of the error estimate due to ill-conditioning of the reconstruction matrices. To mitigate this, we want to assign different importance to the errors over different random subsets. Consider the following singular value decomposition and the resulting expression for the reconstruction error on the set $\setcurve{S}_i^c$:
\begin{equation*}
    \arr{M}_i = \arr{V}_i\greekarr{\Sigma}_i\arr{W}_i^T, \quad\arr{e}(\setcurve{S}_i^c) = \arr{V}_i\greekarr{\Sigma}_i\arr{W}_i^T\greekarr{\beta}.
\end{equation*}
Since $\arr{V}_i$ and $\arr{W}_i$ are orthogonal matrices, the primary scaling of the magnitude of $\greekarr{\beta}$ is due to $\arr{\Sigma}_i$.

To control the scaling of magnitude due to $\greekarr{\Sigma}_i$ we would like to replace the singular values $\sigma$ in $\greekarr{\Sigma}_i$ with $\sigma$, if $\sigma < 1$, and $1$, if $\sigma \geq 1$. This is essentially an operation that clips the singular values to $1$ from above. Although we decomposed $\arr{M}_i$, it is worth keeping in mind that we only have access to $\arr{e}(\setcurve{S}_i^c)$, and in order to control the magnitude of this error,
we can pre-multiply with a matrix.
To achieve the transformation in the singular values, we multiply $\arr{e}(\setcurve{S}_i^c)$ as follows to get a new error
\begin{equation}
    \arr{e}_\text{new}(\setcurve{S}_i^c) = \greekarr{\Sigma}_i'\arr{V}_i^T \arr{e}(\setcurve{S}_i^c),
    \label{eq:err_weight}
\end{equation}
where $\greekarr{\Sigma}_i'$ is a $\abs{\setcurve{S}_i^c}\times\abs{\setcurve{S}_i^c}$ diagonal matrix  with  diagonal entries $1$, if $\sigma < 1$, and $1/\sigma$, if $\sigma \geq 1$. The subset specific weighting in \eqref{eq:err_weight} can be seen as giving more importance to certain vertices while ignoring others. Although the weights in $\greekarr{\Sigma}_i'$ do not directly correspond to the weights on individual vertices, the weights on individual vertices can be seen as combinations of weights on multiplying by the matrix $\arr{V}_i\greekarr{\Sigma}_i'\arr{V}_i^T$. Finally, we estimate the error using
\begin{equation}
    \hat{E}_{\setcurve{S}^c} = \frac{\sum_{i\in\{1, \cdots, k\}} \norm{\arr{e}_\text{new}(\setcurve{S}_i^c)}^2}{k}.
    \label{eq:proposed_estimation}
\end{equation}

\section{Experiments}

\subsection{Graph construction}
\label{sec:graph_constructions}
For the initial verification of our error estimation approach, we construct random regular graphs with $1000$ vertices according to the model \texttt{RandomRegular} from \cite{pygsp}. We define noisy bandlimited signals with bandwidths $\{20, 50, 120\}$ and power $1$ and noise power levels $0.1$ and $0.2$ according to the model in \eqref{eq:noiseModel}. We call these graphs and signals as synthetic graphs and signals for our experiments.

For the next experimental validation, we use publicly available climate data from National Oceanic and Atmospheric Administration \cite{noaa} which has been measured by sensors throughout the United States. The sensor data consists of different weather measurements like average daily temperature, daily precipitation along with the latitudes, longitudes, and altitudes of the corresponding sensors.

Using the locations, we construct graphs by connecting the 5 nearest sensor locations to each sensor. The edge weights of the graph are given by $e^{-d^2/2\sigma^2}$ where we experimentally choose $\sigma=50$. We calculate the distance $d$ between the measurement locations using the latitude, longitude, and altitude of the measuring station using $\sqrt{d_f^2 + d_a^2}$. $d_f$ is the flat distance computed using the \texttt{distance} package from \texttt{geopy} library, and $d_a$ is the altitude. While constructing the graph we drop sensors whose measurements are missing, because there is no way to verify our predictions for those sensors. The measurements that we include as signals are day averages measured on $3^{rd}$ Jan 2020, and monthly normals \cite{dunlop2008dictionary}, which are average measurements for January 2010.

\subsection{Set selection}
In Section \ref{sec:prob_form}, we assume that the signal values on a vertex set $\setcurve{S}$ are known. To select this set for the constructed graphs on which we assume signal values are known, we use the AVM algorithm \cite{jayawant2022practical} to sample 200 vertices from each graph, and observe the reconstruction errors on the frequencies $\{10, 20, \cdots, 110\}$. The only exception is the California sensor network graph where we sample $100$ vertices and observe the reconstruction errors over the frequencies $\{10, 20, \cdots, 80\}$, because the graph itself contains only 300 vertices.

To estimate the reconstruction error using cross-validation,
we partition each sampling set $\setcurve{S}$ into 10 subsets using \texttt{RepeatedKFold} function from \texttt{model\_selection} package of \texttt{sklearn}. We measure the squared reconstruction error on each subset of the partition repeated 50 times, and average over the squared reconstruction errors as per \eqref{eq:proposed_estimation}.

\subsection{Results}
We can see the results of our estimation in Fig. \ref{fig:mainRecon}. The estimated cross-validation error tracks the actual error in the wide variety of the graphs and graph signals that we experiment with. We note that in Fig. \ref{fig:bw20npp2} the actual error increases slightly, however the estimated error does not increase with it. This is due to the error weighting strategy proposed in \eqref{eq:err_weight}. Since for the problem of choosing the bandwidth we are interested in correctly locating the lowest value of the actual error,  the ability of the error estimate to track the actual error as it increases should be of lesser importance than its ability to track the actual error as it decreases. A more accurate error estimation could be achieved with different set selection or error weighting strategies for cross-validation which we reserve for future work.

\section{Conclusion}
In this paper, we proposed a way to minimize graph signal reconstruction error without assuming the knowledge of the signal bandwidth. In the process, we tailored the cross-validation method for the problem of reconstruction error estimation. Our technique estimated the error accurately as a function of the signal bandwidth on a variety of bandlimited signals with noise and also for sensor networks measuring weather.

\newpage
\bibliographystyle{IEEEbib}
\bibliography{cvrefsarxiv}

\begin{thebibliography}{10}

\bibitem{ortega2021introduction}
Antonio Ortega,
\newblock {\em Introduction to Graph Signal Processing},
\newblock Cambridge University Press, 2022.

\bibitem{lopes2005inferential}
Vitor~V Lopes and Jos{\'e}~C Menezes,
\newblock ``Inferential sensor design in the presence of missing data: a case
  study,''
\newblock {\em Chemometrics and intelligent laboratory systems}, vol. 78, no.
  1-2, pp. 1--10, 2005.

\bibitem{guerreiro2002factorization}
Rui~FC Guerreiro and Pedro~MQ Aguiar,
\newblock ``Factorization with missing data for 3d structure recovery,''
\newblock in {\em 2002 IEEE Workshop on Multimedia Signal Processing.} IEEE,
  2002, pp. 105--108.

\bibitem{mott1994climate}
P~Mott, TW~Sammis, and GM~Southward,
\newblock ``Climate data estimation using climate information from surrounding
  climate stations,''
\newblock {\em Applied Engineering in Agriculture}, vol. 10, no. 1, pp. 41--44,
  1994.

\bibitem{chen2015discrete}
Siheng Chen, Rohan Varma, Aliaksei Sandryhaila, and Jelena Kova{\v{c}}evi{\'c},
\newblock ``Discrete signal processing on graphs: Sampling theory,''
\newblock {\em IEEE Transactions on Signal Processing}, vol. 63, no. 24, pp.
  6510--6523, 2015.

\bibitem{tanaka2020sampling}
Yuichi Tanaka, Yonina~C Eldar, Antonio Ortega, and Gene Cheung,
\newblock ``Sampling signals on graphs: From theory to applications,''
\newblock {\em IEEE Signal Processing Magazine}, vol. 37, no. 6, pp. 14--30,
  2020.

\bibitem{anis2015efficient}
Aamir Anis, Akshay Gadde, and Antonio Ortega,
\newblock ``Efficient sampling set selection for bandlimited graph signals
  using graph spectral proxies,''
\newblock {\em IEEE Transactions on Signal Processing}, vol. 64, no. 14, pp.
  3775--3789, 2015.

\bibitem{jayawant2022practical}
Ajinkya Jayawant and Antonio Ortega,
\newblock ``Practical graph signal sampling with log-linear size scaling,''
\newblock {\em Signal Processing}, vol. 194, pp. 108436, 2022.

\bibitem{puy2016random}
Gilles Puy, Nicolas Tremblay, R{\'e}mi Gribonval, and Pierre Vandergheynst,
\newblock ``Random sampling of bandlimited signals on graphs,''
\newblock {\em Applied and Computational Harmonic Analysis}, 2016.

\bibitem{shuman2013emerging}
David~I Shuman, Sunil~K Narang, Pascal Frossard, Antonio Ortega, and Pierre
  Vandergheynst,
\newblock ``The emerging field of signal processing on graphs: Extending
  high-dimensional data analysis to networks and other irregular domains,''
\newblock {\em IEEE Signal Processing Magazine}, vol. 30, no. 3, pp. 83--98,
  2013.

\bibitem{pesenson2008sampling}
Isaac Pesenson,
\newblock ``Sampling in paley-wiener spaces on combinatorial graphs,''
\newblock {\em Transactions of the American Mathematical Society}, vol. 360,
  no. 10, pp. 5603--5627, 2008.

\bibitem{wang2015local}
Xiaohan Wang, Pengfei Liu, and Yuantao Gu,
\newblock ``Local-set-based graph signal reconstruction,''
\newblock {\em IEEE transactions on signal processing}, vol. 63, no. 9, pp.
  2432--2444, 2015.

\bibitem{shao1993linear}
Jun Shao,
\newblock ``Linear model selection by cross-validation,''
\newblock {\em Journal of the American statistical Association}, vol. 88, no.
  422, pp. 486--494, 1993.

\bibitem{pygsp}
Micha\"el Defferrard, Lionel Martin, Rodrigo Pena, and Nathana\"el Perraudin,
\newblock ``Pygsp: Graph signal processing in python,'' .

\bibitem{noaa}
Russell~S. Vose, Scott Applequist, Mike Squires, Imke Durre, Matthew~J. Menne,
  Claude~N. Williams, Chris Fenimore, Karin Gleason, and Derek Arndt,
\newblock ``Improved historical temperature and precipitation time series for
  u.s. climate divisions,''
\newblock {\em Journal of Applied Meteorology and Climatology}, vol. 53, no. 5,
  pp. 1232 -- 1251, 2014.

\bibitem{dunlop2008dictionary}
Storm Dunlop,
\newblock {\em A dictionary of weather},
\newblock OUP Oxford, 2008.

\end{thebibliography}

\end{document}